\documentstyle[12pt,aasms4]{article}
\def\ea{{\it et al. }}
\def\ltsima{$\; \buildrel < \over \sim \;$}
\def\simlt{\lower.5ex\hbox{\ltsima}}            
\def\gtsima{$\; \buildrel > \over \sim \;$}
\def\simgt{\lower.5ex\hbox{\gtsima}}            
\received{~~}
\accepted{~~}
\journalid{}{}
\articleid{}{}
 
\begin{document}

\title{\bf Monitoring Ly$\alpha$ Emission from the Blazar 3C~279}

\author{Anuradha Koratkar, Elena Pian, C. Megan Urry, \and Joseph E. Pesce}
\affil{Space Telescope Science Institute, 3700 San Martin Dr., Baltimore, MD 21218 \\
(koratkar@stsci.edu, pian@stsci.edu, cmu@stsci.edu, pesce@stsci.edu)}

\begin{abstract}

The blazar 3C~279 is well studied and shows frequent large continuum
flares from radio to $\gamma$-ray wavelengths. There have been a number
of multi-wavelength observations of 3C~279, and hence there are many
ultraviolet data for this object available in the UV archives.
In this paper we present Ly$\alpha$ emission line measurements for
3C~279 using all the archival {\it IUE} SWP spectra from 1988 to 1996
and all archival {\it HST/FOS} G190H spectra from 1992 to 1996.

Individual archival {\it IUE} spectra of 3C~279 show weak Ly$\alpha$
emission at $\sim$1868~\AA\ ($z = 0.536$), which is easily seen in the
co-added data. The Ly$\alpha$ emission is observed in all the {\it
HST/FOS} spectra. The strength of Ly$\alpha$ is nearly constant
($\sim$5 $\times 10^{-14}$ erg~cm$^{-2}$ s$^{-1}$), while the
1750~\AA\ continuum varies by a factor of $\sim$50, from $\sim 0.6$ to
31.6 $\times 10^{-15}$ erg~cm$^{-2}$ s$^{-1}$ \AA$^{-1}$.

The behavior of the Ly$\alpha$ emission line flux and continuum flux is
similar to that of the only other well observed blazar, 3C~273, which
shows constant line flux while the continuum varies by a factor of
$\sim$3. This near-constancy of emission line flux in the two
best-studied blazars, suggests that the highly variable beamed
continuum is not a significant source of photoionization for the gas.
Some other source, such as thermal emission from an accretion disk,
must be providing a significant fraction of the photoionizing flux in
these objects. The large amplitude variability seen at $\gamma$-ray
energies must be due to changes in the energetic electrons in the jet
rather than changes in the external photon field.

{\underline{\em Subject Headings:}} galaxies:active --
quasars:individual (3C 279) -- ultraviolet:observation -- variability

\end{abstract}

\section{Introduction} 

Blazars as a class of Active Galactic Nuclei (AGN) exhibit a smooth,
highly variable, highly polarized IR-UV continuum. According to
``Unified Schemes'' for radio-loud AGN (Browne \& Murphy 1987; Barthel
1989; Antonucci 1993) the observed continuum comprises a mixture of
thermal radiation and non-thermal beamed radiation. The thermal,
unbeamed component extending from the UV to the soft X-rays may
originate from a hot accretion disk, as in radio-quiet AGN. The
non-thermal component extending from the radio to the $\gamma$-rays is
thought to be relativistically beamed synchrotron radiation, possibly
originating from a twin jet (K\"onigl 1981). This beamed component is
highly variable and highly polarized. Blazars are thought to be those
objects viewed at a small angle to the beaming axis and thus dominated
by the non-thermal component (Urry \& Padovani 1995).  Some blazars
exhibit a broad emission line spectrum similar to Seyfert galaxies and
radio quiet quasars, particularly when the continuum is faint (the
exceptions are classified as BL Lac objects; see however Scarpa \&
Falomo 1997).  Although blazars are highly variable and show broad
emission lines, their line variability studies have received relatively
little attention to date.

In recent years the study of broad-emission-line variability has been
recognized as a powerful diagnostic of the broad line region in
Seyfert~1 galaxies (Peterson 1993). In the best monitored AGN, high
ionization lines like Ly$\alpha$ and C~IV are more rapidly variable
(factors of $\sim$10) than low ionization lines like Mg II or H$\alpha$
(e.g., Clavel et al. 1991; Peterson 1993). Hence, the high ionization
lines are extremely powerful probes of the smallest, most variable
regions.  Yet Ly$\alpha$ emission line variability has been
investigated in only a few high-luminosity radio-loud objects (Bregman
et al. 1986; P\'erez et al. 1989; Gondhalekar 1990; Ulrich,
Courvoisier, \& Wamsteker 1993; Webb et al.  1994; Scarpa, Falomo, \&
Pian 1995, Wamsteker \ea 1997). Most of these studies indicate that
emission line variability timescales are on the order of months and the
variability is confined to the line cores. However, these objects are
either not typical blazars (i.e., their continuum is not dominated by a
beamed component) or have only a handful of observations.

It is not clear whether the broad emission lines in blazars behave
differently from the broad emission lines observed in Seyfert~1
galaxies. Further, what is the role of the blazar jet in photoionizing
the emission lines; that is, what is the effect of the beamed blazar
light on the line emitting region? What is the structure of the broad
line region gas?  Some investigations are starting to address these
questions (e.g., Corbett et al. 1996).

3C~279 is an excellent target to study emission line variability in
blazars, since it is a highly variable, bright radio source. It is also
one of the brightest known extragalactic $\gamma$-ray sources (Hartman
et al.  1992; von Montigny et al. 1995), is highly variable at X-ray
and $\gamma$-ray energies (Makino et al.  1989; Kniffen et al. 1993;
Maraschi et al. 1994), and is one of the clearest cases of relativistic
beaming (Dondi \& Ghisellini 1995). It shows highly variable optical
polarization (9--45\%; Angel \& Stockman 1980; Impey \& Tapia 1990;
Mead et al. 1990) and has been one of the most intensively monitored
objects of the blazar class. Finally, it is one of two emission-line
blazars with extensive archival UV spectra. In this paper we report the
Ly$\alpha$ emission line history for 3C~279, which has not been as
thoroughly investigated as the continuum, and its relation to the
continuum variability. The analysis is based on archival ultraviolet
data from both the {\it International Ultraviolet Explorer} satellite
({\it IUE}) and the {\it Hubble Space Telescope} ({\it HST}).

\section{Observations}
\subsection{HST/FOS Data}

Between 1992 and 1996, 3C 279 was observed three times with the {\it
HST} Faint Object Spectrograph (FOS). The 1992 observation covers the
wavelength range 1150 \AA\ -- 4750 \AA, and has been discussed in great
detail by Netzer et al.  (1994) who note that the spectrum shows the
Ly$\alpha$, C~IV $\lambda$1549, C~III] $\lambda$1909, Mg~II
$\lambda$2800, and H$\beta$ lines. The Ly$\alpha$, C~III], and H$\beta$
line profiles are fairly symmetric, while C~IV and Mg~II have
substantial red wings. During this period 3C~279 was in a faint state,
with a clearly visible Ly$\alpha$ emission line (see Figure~1).  The
1994 and 1996 observations cover the wavelength range 1150 \AA\ -- 2312
\AA. Both these observations were obtained when 3C 279 showed a flare.
Once again, in these FOS spectra the Ly$\alpha$ is still easily visible
(see Figure~1), but the signal-to-noise ratio (S/N) is not as good as
in the 1992 observation.

We retrieved the {\it FOS} data from the {\it HST} archive and
recalibrated them using the best reference files, as recommended by the
{\it HST} Data Handbook (Version 2).  The details of each observation
are given in Table~1 and the spectra are shown in Figure~1.  The
recalibration of the {\it FOS} spectra was necessary because the older
spectra are calibrated using different procedures and could not
otherwise be directly compared to more recent spectra.  The
recalibrated {\it FOS} spectra can also be directly compared with the
{\it IUE} spectra, since both the {\it HST} and {\it IUE}
spectro-photometric calibrations use the model white dwarf reference
scale (Colina and Bohlin 1994). In fact, based on an analysis of
Seyfert galaxies observed simultaneously with {\it FOS} and {\it IUE},
absolute flux differences are $\simlt 5$\% (Koratkar et al. 1997).

The 1994 FOS observation was obtained using two gratings (G130H and
G190H). During the latter observation, due to reacquisition problems,
the target was partially out of the aperture and the G190H spectrum was
affected. We have therefore added a constant flux to the G190H
observation to align the data with the G130H spectrum. The errors given
in Table 1 for this spectrum (y2ey0606) reflect our uncertainty (10\%)
in applying this procedure.

\subsection{IUE Data}
 
Between 1988 and 1996, 3C 279 was observed with {\it IUE} many times
with both the short-wavelength and long-wavelength cameras. The long
wavelength (LWP/LWR) spectra are affected by low detector sensitivity,
and, increasingly with time, by scattered light, hence emission lines
longward of the Ly$\alpha$ line are difficult to detect. In this paper,
we concentrate on the Ly$\alpha$ line, and thus only the 15
short-wavelength (SWP) camera observations are discussed further.

A careful inspection of the 15 line--by--line spectral images for flaws
showed that in one case (SWP 40489), the Ly$\alpha$ spectral region was
contaminated by a cosmic ray. This spectrum was therefore rejected from
further analysis.

All the line--by--line spectral images were also analyzed to determine
the background contribution. A comparison of the average Flux Numbers
(FN) in the 1866-1872 \AA\/ interval centered on the object spectrum
and on the background (10 lines away on either side of the object
spectrum) showed that in SWP 46662 and SWP 53261 the background
contribution exceeded 90\% of the signal+background emission. Therefore
the  Ly$\alpha$ emission line flux measurements for these spectra in
Table 1 are 5$\sigma$ upper limits.  These noisy spectra have not been
used in the co-addition described below.

All the spectra were extracted from the line--by--line image files
using the NEWSIPS technique being implemented in the final {\it IUE}
archive.  A summary of the NEWSIPS extraction technique can be found in
Nichols \& Linsky (1996). The details of each observation are given in
Table~1.  All the 1992-1993 spectra are under exposed, but some signal
is present.  {\it IUE} spectral regions with known artifacts, at
1278-1279 \AA , 1450-1550 \AA , and 1663 \AA\ (Crenshaw, Norman, \&
Bruegman 1990), were excluded from the analysis.

The improved S/N obtained through NEWSIPS allowed sufficient contrast
in the SWP spectra between the continuum and Ly$\alpha$, even in the
lowest states, to perform for the first time a systematic study of the
emission line in this source. The Ly$\alpha$ emission feature that is
clearly present in the {\it FOS} data is also seen in all but four
individual SWP spectra (see Table~1). Recently, the sensitivity
degradation correction for SWP spectra obtained after 1993 has been
determined. This correction is time dependent, and can be up to 5\% in
1996 (Imhoff 1997). This correction has not yet been implemented in the
final {\it IUE} archive, however, so we have conservatively added a 5\%
error in quadrature to the uncertainties of the continuum and line
fluxes for spectra obtained after 1993.

To improve the S/N ratio of the Ly$\alpha$ emission line, we binned the
spectra in four groups depending on the continuum flux at 1750 \AA. The
SWP spectra in each group were then co-added, as follows:  the `high
spectrum' was generated by co-adding all spectra with $f_{cont} >$ $2
\times 10^{-14}$ erg cm$^{-2}$ s$^{-1}$ \AA$^{-1}$, the `medium
spectrum' is a co-addition of all spectra with $8\times10^{-15} <
f_{cont} \leq 2\times10^{-14}$ erg cm$^{-2}$ s$^{-1}$ \AA$^{-1}$, the
`low spectrum' is a co-addition of all spectra with $3\times10^{-15} <
f_{cont} \leq 8\times10^{-15}$ erg cm$^{-2}$ s$^{-1}$ \AA$^{-1}$, and
the `very low spectrum' includes all spectra with $1\times10^{-15} \leq
f_{cont} \leq 3\times10^{-15}$ erg cm$^{-2}$ s$^{-1}$ \AA$^{-1}$.  In
Figure~2 we show the four co-added spectra.  The Ly$\alpha$ emission
line is easily seen at $\sim$ 1868 \AA.  The redshift measured from the
{\it IUE} observations is in good agreement with the redshift
determined from the high S/N ratio {\it FOS} data (Netzer et al. 1994)
given the wavelength uncertainty and the low  S/N ratio of the {\it
IUE} spectra.

\subsection{Flux Measurements}

The Galactic hydrogen column density in the direction of 3C 279 is
2.22$\times$10$^{20}$cm$^{-2}$ (Elvis, Lockman, \& Wilkes 1989) which
implies A$_V$ = 0.13 for a total-to-selective extinction ratio R =
3.1.  All the {\it FOS} and {\it IUE} spectra were dereddened assuming
the Cardelli, Clayton, \& Mathis (1989) extinction curve.

A simple power-law model was fitted to the dereddened spectra in the
wavelength region 1230 -- 1950 \AA\ using the IRAF/STSDAS tool {\it
nfit1d}.  Wavelength regions affected by the {\it IUE} artifacts were
excluded from the fits to the {\it IUE} spectra, as was the region
around Ly$\alpha$.  We defined 1750 \AA\ as the fiducial wavelength for
continuum flux measurements because this region of the SWP spectrum is
relatively free of {\it IUE} artifacts and line emission.

After subtracting the best-fit power-law continuum, the Ly$\alpha$ line
was measured as follows. The 1992 {\it FOS} observation, which has the
best S/N and spectral resolution, was accurately modeled with a
multicomponent Gaussian fit using a minimalist approach (least number
of free parameters).  IRAF/STSDAS tool {\it ngaussfit} was used for
this procedure. This model of the Ly$\alpha$ line was then used to fit
the remaining two {\it FOS} spectra. Since these spectra have lower
S/N, the observed wavelength of the Ly$\alpha$ line was fixed at
1868\AA\ (within the wavelength calibration error of the {\it FOS}) and
the widths and amplitudes were allowed to vary.  We note here that all
the {\it FOS} spectra show an intrinsic absorption feature in the
Ly$\alpha$ emission line.  These absorption features are discussed in
detail by Chaffee et al. (1997).

Due to the low S/N ratio of the {\it IUE} data, the observed wavelength
of the Ly$\alpha$ line was fixed at 1868~\AA\ (within the wavelength
calibration error of the {\it IUE}), and the line was modeled with a
single Gaussian whose width and height were allowed to vary freely.  A
single Gaussian fit was sufficient for the {\it IUE} because of the low
spectral resolution. In all but two spectra, the Ly$\alpha$ emission
line is easily detected; in the exceptions (SWP35443, SWP36420) the
Ly$\alpha$ region is very noisy.

The continuum fluxes at 1750 \AA\/ and the Ly$\alpha$ line fluxes,
along with their associated uncertainties, are reported in Table~1 for
the individual spectra and in Table~2 for the co-added spectra.

\section{Results and Discussion} 

Although the data are sparse and unevenly sampled, the continuum flux
density and the line flux over the 8-year observation period show very
different characteristics (Figures 3 and 4).  The continuum flux has
varied by a factor of $\sim$50, while the Ly$\alpha$ line flux has
remained nearly constant. We used two independent methods to test
whether the emission line flux is constant.  In the first method, a
reduced chi-square test on the emission line flux gives a $\chi^2$ =
1.3 for 12 degrees of freedom indicating that the emission line flux
data are consistent with being constant. In the second method, in our
line fitting procedure (discussed in section 2) we kept all parameters
of the Gaussian model constant and determined the reduced chi-square of
the fit for each spectrum.  The constant model used the parameters
derived from the 1992 {\it FOS} spectrum, which has the best S/N and
spectral resolution.  The reduced chi-square of the constant model and
the free Gaussian model were checked using an F-test.  This test showed
that the constant model and the free Gaussian model were comparable at
the 95\% level (F=1.1 for a change of 2 degrees of freedom).  We
conclude that in 3C~279, although the observed continuum flux density
has varied, the Ly$\alpha$ flux has remained essentially constant.

This trend between the continuum and line is similar to the case of
3C~273, where the Ly$\alpha$ varied little or not at all during a
factor of three change in the UV continuum (Ulrich et al. 1993). Such a
trend between the emission line fluxes and the continuum was first
noted by Sandage, Westphal and Strittmatter (1966).  The continuum and
line variation in 3C~279 is in sharp contrast to that seen in Seyfert
galaxies, where the continuum and line vary with similar amplitudes and
in a correlated way (Koratkar \& Gaskell 1991, Clavel \ea 1991,
Peterson 1993, Reichert \ea 1994, Crenshaw \ea 1996 and Wanders \ea
1997). This lack of Ly$\alpha$ variation is not just an effect of
temporal sampling of the variability. In order to determine the effect
of temporal sampling on the line variability, we compared our results
to those of the much better-understood case of the Seyfert galaxy NGC
5548 which shows a clear correlation ($r=0.6$, $p\sim99\%$) between
Ly$\alpha$ and the UV continuum (Clavel et al. 1991; Korista et al.
1995), with a lag of $\sim$10 days.  In 1989, NGC~5548 was observed
every four days for eight months (Clavel et al. 1991).  We selected 16
points at random from the 1989 NGC~5548 continuum and Ly$\alpha$ light
curves. For these 16 points we determined the amount of variation in
both the continuum and lines flux. The process was repeated a 100
times. There was not a single 16 point sampling where the line flux was
constant while the continuum flux varied as seen in 3C~279.  This
difference in the Ly$\alpha$ variability of blazars can be explained if
the thermal continuum that primarily photoionizes the line emitting gas
is hidden by the variability of the beamed continuum in radio-loud
AGN.  This effect would be strongest in blazars, where the beamed
component contribution dominates the observed continuum.

The Ly$\alpha$ equivalent width ($W_{Ly\alpha}$) of 3C~279 ranges
between 1\AA\/ and 45\AA\/ (Figure 5).  These values of $W_{Ly\alpha}$
are smaller than those observed in normal quasars (Kinney, Rivolo \&
Koratkar 1990; Cristiani \& Vio 1990; Cheng \& Fang 1987), where 
the thermal component, possibly the accretion disk, is the only source
of ionizing radiation and observed continuum and gives rise to the
observed $W_{Ly\alpha} \sim$100\AA\/.  In blazars there are two sources
of ionizing radiation: (1) the beamed non-thermal radiation from the
relativistic jet, and (2) the thermal component probably from an
accretion disk. Thus, the observed continuum has contributions both
from the disk and jet. If we assume that the beam is not contributing
to ionizing the line emitting gas, and only the thermal component is
responsible for line emission, then the contribution of the beamed
component to the observed continuum determines the strength of the
equivalent width.  The expected $W_{Ly\alpha}$ in blazars should be
smaller than observed in normal quasars and should be anti-correlated with the
contribution of the beamed component to the observed continuum (see
Corbett \ea 1996).  In 3C~279 we see this characteristic equivalent
width trend (see Figure 5).  At low observed continuum levels, the
$W_{Ly\alpha}$ becomes comparable ($\sim$100\AA\/) to other quasars
when the beamed continuum contribution is $\sim$2$\times$ the accretion
disk contribution, and at high observed continuum levels the beamed
continuum contribution is $\sim$30$\times$ the accretion disk
contribution.  This model suggests that at low continuum levels we
should see a change in UV continuum slope because of the increased
contribution of the big blue bump due to the accretion disk. At high
continuum levels we should just see the synchrotron power law slope.
Such a change in the UV slope of 3C~279 has been noted by Pian \ea
(1997).

One of the models for the production of $\gamma$-ray emission in
blazars is Compton upscattering of ambient Ly$\alpha$ line photons
(Sikora, Begelman, \& Rees 1994).  Variability of $\gamma$-rays could
therefore be caused by changes in the Ly$\alpha$ intensity or in the
energetic electrons of the jet. The $\gamma$-rays would vary linearly
with Ly$\alpha$ flux and/or quadratically with respect to synchrotron
emission depending on the direction of the jet electrons.  If some of
the Ly$\alpha$ flux increases in response to the beamed continuum
(Ghisellini \& Madau 1996), then the $\gamma$-ray flux could increase
even more than the square of the observed UV synchrotron flux.  A
strong correlation between UV and $\gamma$-ray flux has been observed
in 3C~279 (Maraschi \ea 1994, Wehrle \ea 1997), indicating that the
beamed continuum could be ionizing some of the line emitting gas.  Our
present analysis suggests that $\gamma$-ray variability must be caused
by variations in the jet rather than the external Ly$\alpha$ photons.
The present data do not allow us to determine the contribution of jet
synchrotron emission to photoionizing some of the Ly$\alpha$ emission
clouds.  If the beamed continuum does ionize some of the line emitting
gas, we should see rapid variations in the emission line profile. The
nature of the variations would depend on the kinematic structure of the
line emitting gas. Our {\it IUE} data do not have high enough spectral
resolution or S/N to allow us to decompose the line core and wings but
the Ly$\alpha$ line profile appears symmetric, as it does in the {\it
FOS} 1992 spectrum (Netzer et al. 1994). The 1994 and 1996 {\it FOS}
spectra show intrinsic absorption and do not have sufficient S/N to
determine accurately the line profile variations.  Simultaneous
monitoring of 3C~279 in both the UV and $\gamma$-ray would allow us to
study continuum and emission line profile variability, and then
unambiguously select the correct inverse Compton model for $\gamma$-ray
production.

\section{Conclusions}

Our analysis of the archival {\it IUE} and {\it HST/FOS} observations
of 3C~279 over a period of eight years shows that although the
continuum has varied by large amplitude, the Ly$\alpha$ flux has
remained essentially constant.  This suggests that the beamed continuum
is not the dominant source of photoionization of the broad line clouds,
which are instead likely to be powered by a thermal component
underlying the beamed continuum.

The strong $\gamma$-ray emission from 3C~279 may be due to Compton
scattering of the Ly$\alpha$ line photons. In this case, the large
amplitude of the $\gamma$-ray variability must be due to changes in the
energy distribution of electrons in the jet, rather than changes in the
external photon field. This implies that $\gamma$-rays will change as
the square of the synchrotron flux, or perhaps even more if the beamed
continuum ionizes line-emitting gas lying close to the jet axis
(Ghisellini \& Madau 1996). High quality line profile observations are
necessary to test this model further.

\acknowledgements

J.E.P., E.P., and C.M.U., would like to acknowledge support from NASA
grants NAG5-1918, NAG5-1034, and NAG5-2499. E.P. acknowledges a
NATO-CNR Advanced Fellowship. We thank C. Imhoff for providing the
NEWSIPS reprocessed {\it IUE} data and for answering numerous
questions. We thank Gabriele Ghisellini and Stefan Wagner for providing
the 1996 SWP spectrum while still proprietary.  This research has made
use of the NASA/IPAC Extragalactic Database (NED) which is operated by
the Jet Propulsion Laboratory, California Institute of Technology,
under contract with the National Aeronautics and Space Administration,
and of NASA's Astrophysics Data System Abstract Service (ADS).

\vskip -1.0in
\begin{table}
\begin{center}
\small
\begin{tabular}{lcrcc}
\multicolumn{5}{c}{{\bf Table 1:} Fits to the Individual Spectra} \\ 
\tableline\tableline
~~Image/ & Date & Exp. time & $F_{\lambda_{1750}}$ & $F_{Ly\alpha}$ \\ 
Spectrum & dd/mm/yy& (sec)~~~ & ($\times 10^{-15}$ erg cm$^{-2}$ s$^{-1}$\AA$^{-1}$) & ($\times 10^{-14}$ erg cm$^{-2}$ s$^{-1}$) \\ \tableline
SWP33864 & 05/07/88 &  6000 & 29.43$\pm$0.36 & 12.17$\pm$4.96 \\ 
SWP33865 & 05/07/88 & 10200 & 31.55$\pm$0.29 &  6.63$\pm$1.95 \\
SWP35443 & 28/01/89 & 10740 & 13.83$\pm$0.21 & 11.20 \tablenotemark{a} \\
SWP36420 & 08/06/89 & 13200 & 17.09$\pm$0.17 & 16.35 \tablenotemark{a} \\
SWP40489 & 29/12/90 & 14400 & 18.34$\pm$0.14 & $\cdots$ \tablenotemark{b} \\ 
SWP42132 & 27/07/91 & 14400 & 11.69$\pm$0.11 &  8.09$\pm$1.60 \\ 
SWP44806 & 29/05/92 & 10800 & 12.24$\pm$0.26 & 11.19$\pm$3.50 \\ 
SWP46649 & 02/01/93 & 13200 &  1.42$\pm$0.12 &  4.52$\pm$1.29 \\
SWP46653 & 03/01/93 & 36000 &  1.47$\pm$0.11 &  3.88$\pm$0.98 \\ 
SWP46657 & 04/01/93 & 39600 &  1.27$\pm$0.10 &  5.33$\pm$0.70 \\ 
SWP46662 & 05/01/93 & 39600 &  1.03$\pm$0.10 &  3.72\tablenotemark{a} \\ 
SWP49681 & 24/12/93 & 12000 &  3.45$\pm$0.21 &  7.38$\pm$1.33 \\ 
SWP49686 & 25/12/93 & 19800 &  4.19$\pm$0.24 &  4.44$\pm$1.15 \\ 
SWP53261 & 03/01/95 & 16800 &  0.60$\pm$0.10 &  5.38 \tablenotemark{a} \\
SWP56635 & 25/01/96 & 19500 &  4.59$\pm$0.31 &  5.27$\pm$1.55 \\
y0pe0b02/3 & 08/04/92 & 2250 & 1.89$\pm$0.03 &  5.22$\pm$0.56 \\
y2ey0606 & 05/06/94 &  1980 &  6.12$\pm$0.08 &  5.86$\pm$1.41 \\ 
y32j0605 & 08/01/96 &  2250 &  5.71$\pm$0.03 &  4.04$\pm$1.19 \\
\tableline
\end{tabular}
\end{center}
\normalsize
\tablenotetext{\rm a}{~The amplitude of Ly$\alpha$ is less than the rms 
fluctuations about the best-fit power-law model so values given are
5$\sigma$ upper limits. High background seen in some of the spectra.} 
\tablenotetext{\rm b}{~Cosmic ray hit exactly on the Ly$\alpha$ line, flux
not measured.}
\end{table}

\clearpage

\vskip -1.0in
\begin{table}
\begin{center}
\begin{tabular}{lccc}
\multicolumn{4}{c}{{\bf Table 2:} Fits to the Co-Added Spectra} \\ 
\tableline\tableline
Image & Co-added spectra & $F_{\lambda_{1750}}$ & $F_{Ly\alpha}$ \\ 
      & & ($\times 10^{-15}$ erg cm$^{-2}$ s$^{-1}$ \AA$^{-1}$) 
      & ($\times 10^{-14}$ erg cm$^{-2}$ s$^{-1}$) \\ 
\tableline
`HIGH' & SWP33864, SWP33865 & 30.66 $\pm$ 0.47 & 10.86 $\pm$ 6.45 \\ 
\tableline
`MEDIUM' & SWP35443, SWP36420 & 13.71 $\pm$ 0.36 & 5.00 $\pm$ 2.45 \\
	& SWP42132, SWP44806 &&\\
\tableline
`LOW'	& SWP49681, SWP49686 & 4.12 $\pm$ 0.14 & 5.26 $\pm$ 2.12 \\ 
	& SWP56635 &&\\ 
\tableline
`VERY LOW' & SWP46649, SWP46653 & 1.40 $\pm$ 0.11 & 4.20 $\pm$ 2.11 \\
	& SWP46657 &&\\ 
\tableline
\end{tabular}
\end{center}
\end{table}
\clearpage

\begin{figure} \caption{The three {\it HST/FOS} spectra of 3C~279. Note
that even in the low state, the Ly$\alpha$ line is clearly visible at
1868 \AA\/. (a) 1992; (b) 1994; (c) 1996.} \end{figure}

\begin{figure} \caption{The co-added {\it IUE} spectra of 3C~279 in
four intensity states, as defined in \S~2.2 (see also Table~2).  In all
continuum flux ranges the Ly$\alpha$ line is visible at 1868 \AA\/. (a)
`High' spectrum; (b) `medium' spectrum; (c) `low' spectrum; (d) `very
low' spectrum.} \end{figure}

\begin{figure} \caption{(a) The continuum flux density (filled points) and
Ly$\alpha$ line flux (open points) of 3C~279 as a function of time.
Over the eight years of observations the source faded by a factor of
$\sim50$, followed by a slight increase at the most recent epoch.  The
{\it IUE} data are represented by circles and the {\it HST/FOS} data by
squares. Error bars are smaller than the symbol if not indicated in the
figure. (b) Same as (a) but for co-added {\it IUE} spectra (pentagons) and 
{\it HST/FOS} data (squares).} \end{figure}

\begin{figure} \caption{(a) The Ly$\alpha$ line flux of 3C~279 as a
function of the continuum flux density (circles, {\it IUE}; squares,
{\it HST/FOS}). (b) The Ly$\alpha$ line flux of 3C~279 as a function of the
continuum flux density for the co-added spectra and the {\it HST/FOS}
(pentagons, {\it IUE} coadded; squares, {\it HST/FOS}) Over a period of 8
years the continuum flux varies by a factor $\sim 50$ while the line
flux remained nearly constant. } \end{figure}

\begin{figure} \caption{The Ly$\alpha$ equivalent width as a function
of the continuum flux density (circles, {\it IUE} individual;
pentagons, {\it IUE} coadded;squares, {\it HST/FOS}). Such a decline in
equivalent width is possible if the line is primarily ionized by the
thermal component, but the flux in the observed continuum has varying
contributions from the beamed continuum. The solid line represents the
variation in equivalent width for a constant line intensity and a varying
beamed continuum.} \end{figure}


\begin{references}

\reference{}Angel, J. R. P., \& Stockman, H. S. 1980, ARAA, 18, 321
\reference{}Antonucci, R. 1993, ARA\&A, 31, 473
\reference{}Barthel, P. D. 1989, ApJ, 336, 606
\reference{}Bregman, J. N., Glassgold, A. E., Huggins, P. J., \& Kinney, A. L.
 1986, ApJ, 301, 698
\reference{}Browne, I. W. A., \& Murphy, D. W. 1987, MNRAS, 226, 601
\reference{}Cardelli, J., Clayton, G.C., \& Mathis, J. S., 1989, ApJ, 345
\reference{}Cheng F., \& Fang, 1987, MNRAS, 226, 485
\reference{}Chaffee, F., et. al. 1998, in preparation
\reference{}Clavel, J., et al. 1991, ApJ, 366, 64
\reference{}Colina, L. \& Bohlin, R.C., 1994, Instrument Science Report
on Standard Calibration Sources CAL/SCS-003
\reference{}Corbett, E. A., Robinson, A., Axon, D. J., Hough, J. H., Jeffries,
 R. D., Thurston, M. R., \& Young, S. 1996, MNRAS, 281, 737
\reference{}Crenshaw, D. M., Norman, D. J., \& Bruegman, O. W. 1990, PASP, 
 102, 463
\reference{}Crenshaw, D. M., \ea 1996, ApJ, 470, 322
\reference{}Cristiani, S., \& Vio 1990, AA, 227, 385 
\reference{}Dondi, L., \& Ghisellini, G. 1995, MNRAS, 273, 583
\reference{}Elvis, M., Lockman, F. J., \& Wilkes, B. J., 1989, AJ, 97, 777
\reference{}Ghisellini, G., \& Madau, P. 1996, MNRAS, in press
\reference{}Gondhalekar, P. M. 1990, MNRAS, 243, 443
\reference{}Hartman, R. C., et al. 1992, ApJ, 385, L1
\reference{}Impey, C. D., \& Tapia, S. 1990, ApJ, 354, 124
\reference{}Kinney, A. L., Rivolo, A., \& Koratkar, A. 1990, ApJ, 357, 338 
\reference{}Kniffen, D. A., et al. 1993, ApJ, 411, 133
\reference{}K\"onigl, A. 1981, ApJ, 243, 700
\reference{}Koratkar, A., et al. 1997, ApJ, submitted
\reference{}Koratkar, A. P., \& Gaskell, C. M., 1991, ApJS, 75, 719
\reference{}Korista, K. T., et al. 1995, ApJS, 97, 285
\reference{}Makino, F., et al. 1989, ApJ, 347, L9
\reference{}Maraschi, L., et al. 1994, ApJ, 435, L91
\reference{}Mead, A. R. G., et al. 1990, A\&AS, 83, 183
\reference{}Netzer, H., et al. 1994, ApJ, 430, 191
\reference{}Nichols, J. S., \& Linsky, J. L. 1996, AJ, 111, 517
\reference{}P\'erez, E., Penston, M. V., \& Moles, M. 1989, MNRAS, 239, 75
\reference{}Peterson, B. M. 1993, PASP, 105, 247
\reference{}Reichert, G. A. \ea 1994, ApJ, 425, 582
\reference{}Sikora, M., Begelman, M. C., \& Rees, M. J. 1994, ApJ, 421, 153
\reference{}Scarpa, R., \& Falomo, R. 1997, A\&A, in press
\reference{}Scarpa, R., Falomo, R., \& Pian, E. 1995, A\&A, 303, 730
\reference{}Ulrich, M.-H., Courvoisier, T. J.-L., \& Wamsteker, W. 1993, 
 ApJ, 411, 125
\reference{} Wehrle, A.E. \ea 1997 in prep.
\reference{}Urry, C. M., \& Padovani, P. 1995, PASP, 107, 803
\reference{}von Montigny, C., et al. 1995, ApJ, 440, 525
\reference{}Wamsteker, W., Wang, T., Schartel, N., \& Vio, R. 1997 MNRAS, submitted
\reference{}Wanders, I., \ea 1997, ApJ, submitted
\reference{}Webb, J. R., et al. 1994, ApJ, 422, 570
\reference{}Zheng, W. 1996, AJ, in press

\end{references}
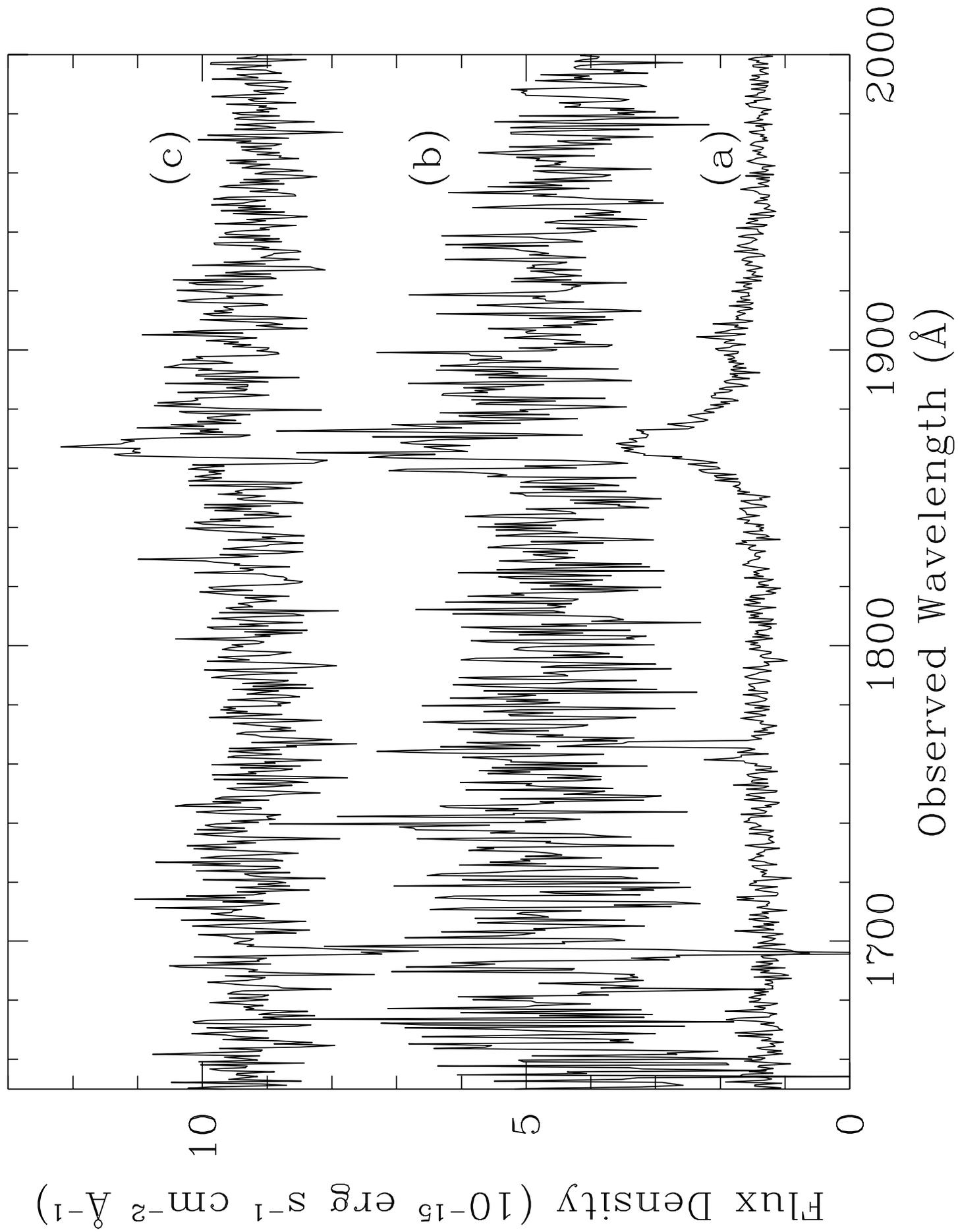
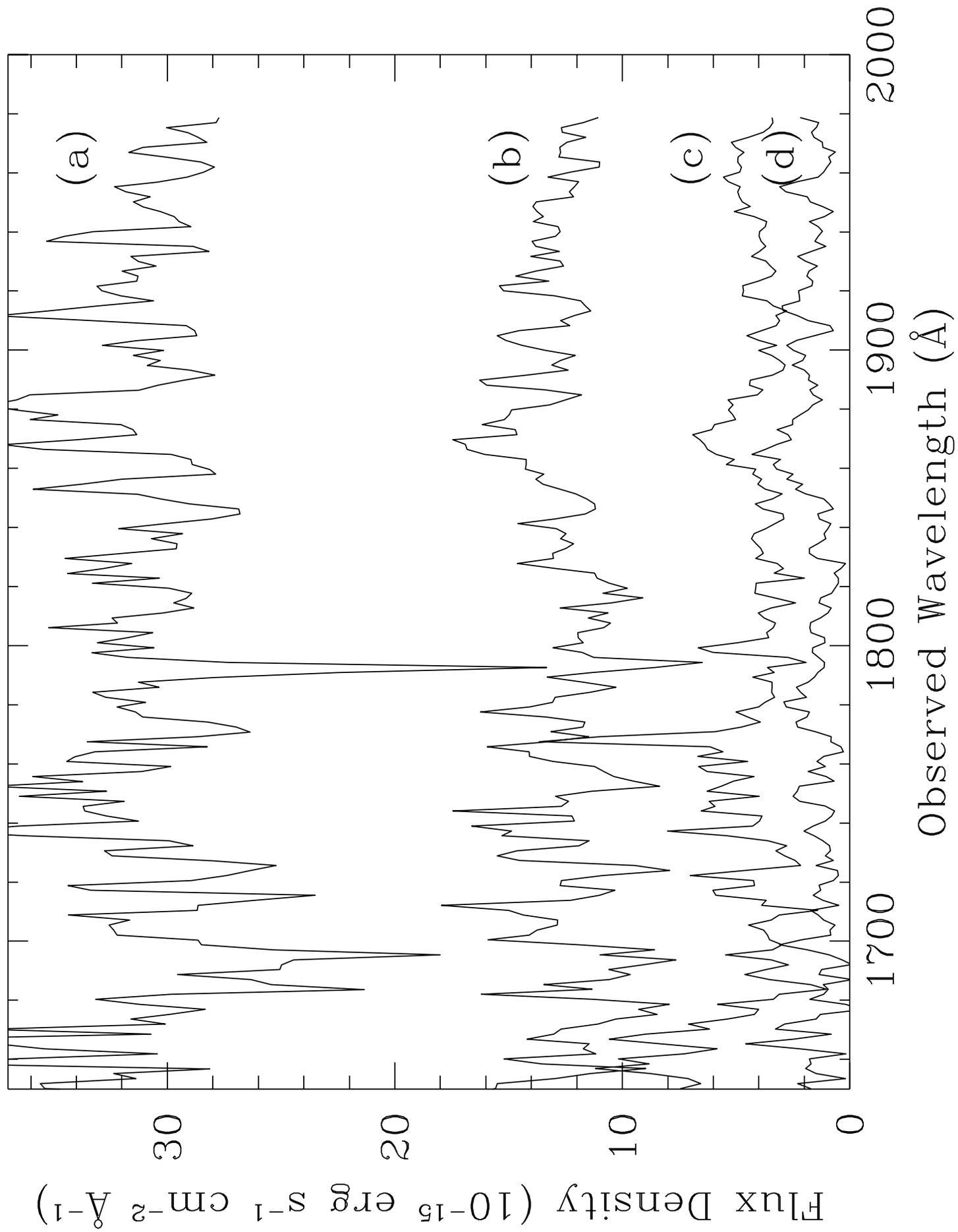
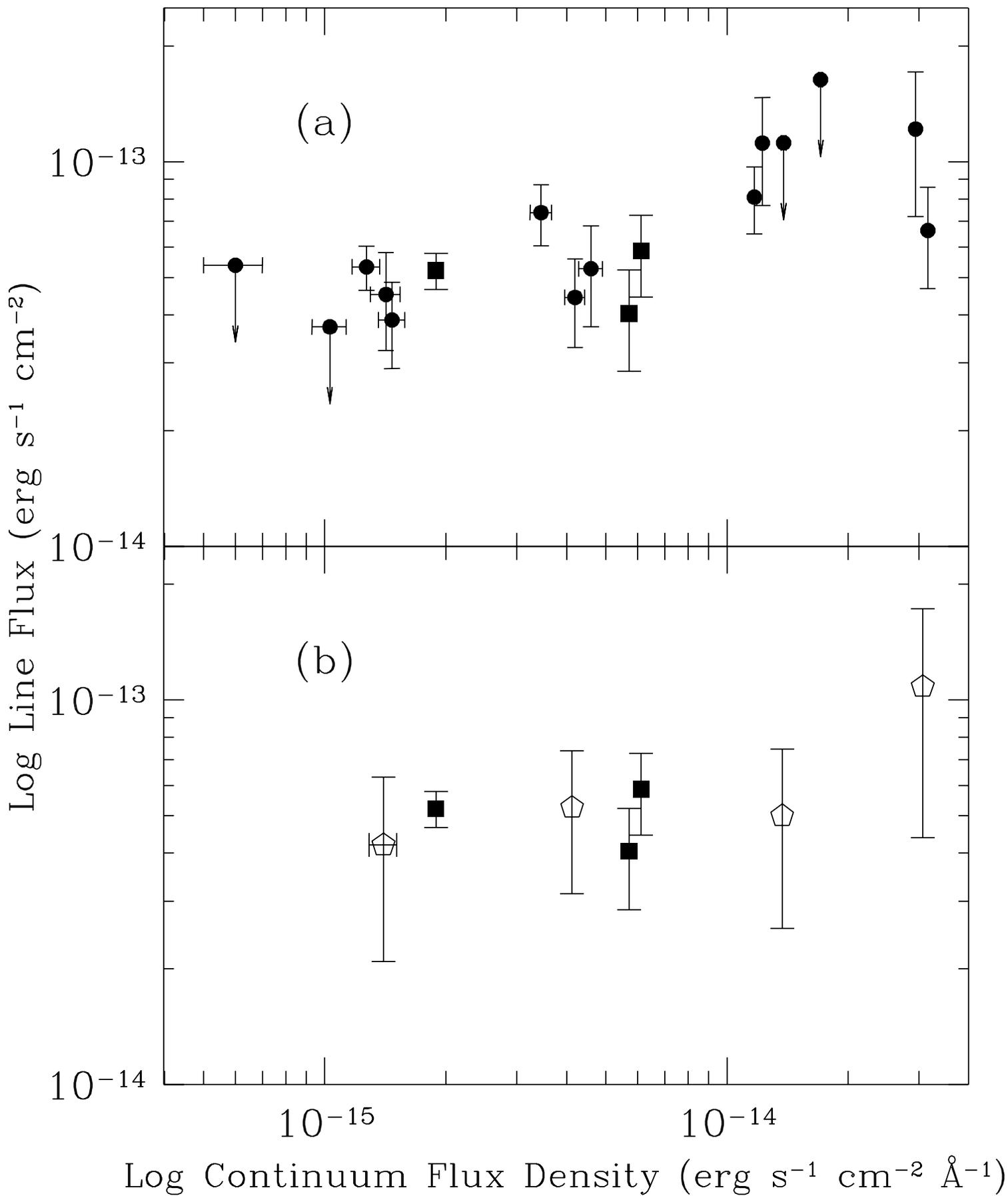
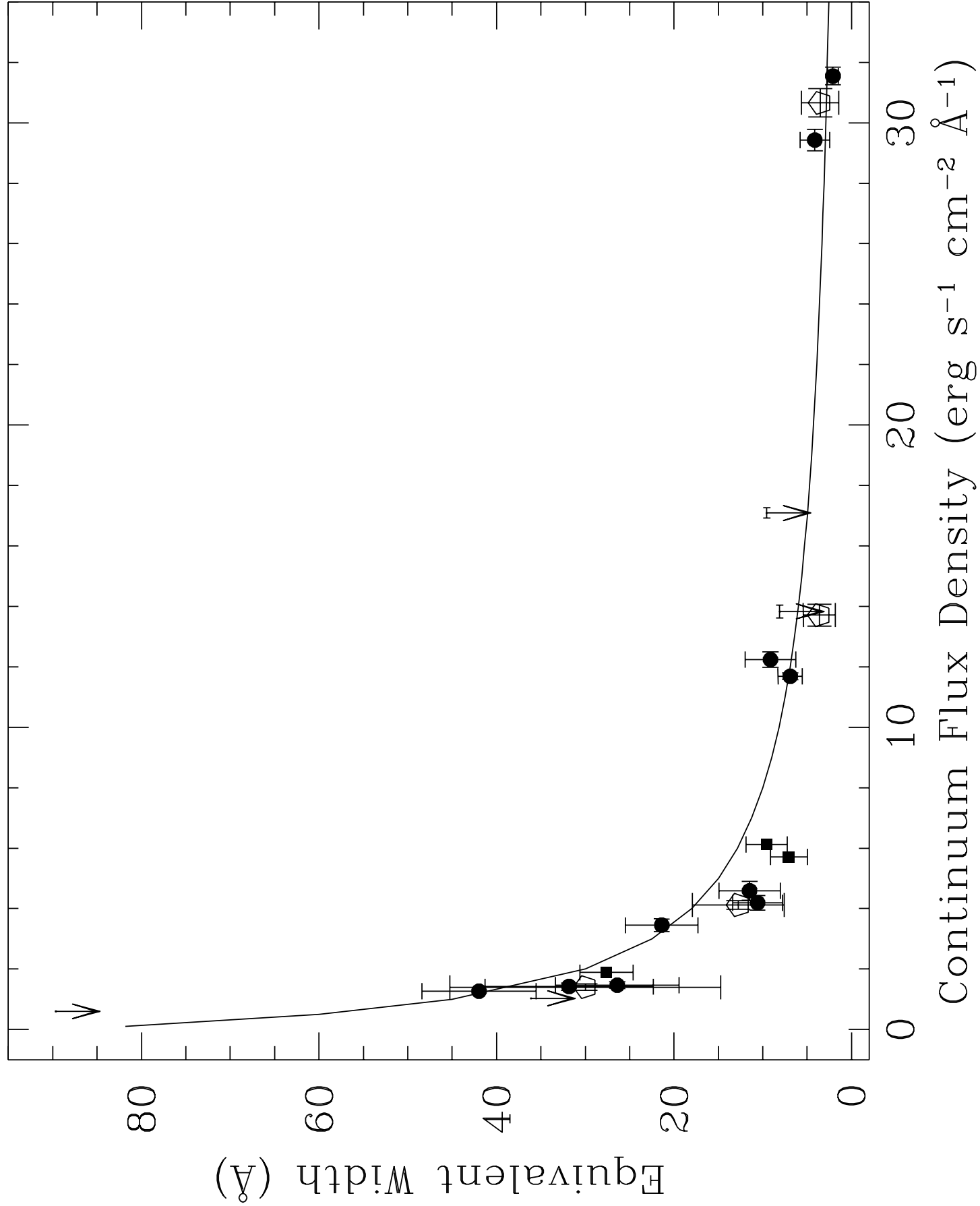
\end{document}